\documentstyle[11pt,twoside,paspconf,epsf]{article}
\def\spose#1{\hbox to 0pt{#1\hss}}
\def\lta{\mathrel{\spose{\lower 3pt\hbox{$\mathchar"218$}}
     \raise 2.0pt\hbox{$\mathchar"13C$}}}
\def\gta{\mathrel{\spose{\lower 3pt\hbox{$\mathchar"218$}}
     \raise 2.0pt\hbox{$\mathchar"13E$}}}

\begin{document}

\title{High velocity gas from the Galactic dark halo}

\author{Mark Walker \& Mark Wardle}

\affil{Special Research Centre for Theoretical Astrophysics, School of Physics,
University of Sydney, NSW 2006, Australia}

\begin{abstract}
{We present the germ of a new model for High Velocity Clouds, derived from
the idea that the dark matter halo of our Galaxy is in the form of cold,
planetary-mass gas clouds. In this picture HVCs arise as a result of
disruptive collisions between dark matter clouds: high velocity atomic
gas is a natural consequence of the dark halo kinematics, and is intimately
associated with assembly of the visible Galaxy. Quasi-spherical halo models
predict a broad 21~cm line background,
together with a number of individually detectable, low-mass HVCs conforming
to a particular velocity field. A  halo model which incorporates satellite
substructure -- after the fashion of the current paradigm of hierarchical
structure formation -- includes both these components and, in addition, some
massive HVCs. These latter HVCs are simply the wakes of the orbiting satellite
halos, and each may have a mass up to 0.3\% of the mass of the satellite.}
\end{abstract}

\section{Introduction}
The suggestion that Galactic dark matter might be in the form of cold
molecular gas clouds was made several years ago (Pfenniger, Combes
\& Martinet 1994). These authors proposed gas in the form of
fractal agglomerates, distributed in a thin disk, but one might equally
well imagine isolated clouds distributed in a quasi-spherical halo
(Gerhard \& Silk 1996). The attractiveness of this picture was greatly
increased by the recognition  (Walker \& Wardle 1998a) that one {\it infers\/}
such a halo from data on ``Extreme Scattering Events'' (Fiedler et al 1987).
The cold clouds are deduced to be molecular, with individual masses
$\sim10^{-3}\;{\rm M_\odot}$; a summary of the relevant astrophysics
will be given elsewhere (Walker \& Wardle 1998b, PASA in preparation).

Another startling piece of evidence in favour of the cold cloud picture
was uncovered by Walker (1998), who showed that this model offers
a simple explanation for the dynamical regularities of spiral galaxies
--- specifically the disk-halo conspiracy and the Tully-Fisher relation.
These regularities arise as a consequence of the conversion of dark matter
into visible (warm, diffuse) gas, via the process of cloud-cloud collisions,
and can be modelled with kinetic theory applied to isothermal halos. This
discovery has served to focus our attention on collisions in connection with
visible galaxy assembly. A basic feature of the collision process is that it
populates the Galaxy with atomic gas moving in orbits which are quite distinct
from the cold rotation of disk gas and so, as we shall see, High Velocity Clouds
are expected. In \S2 we consider collisions occurring within an isothermal,
spherical halo; \S3 generalises this to a halo which has some rotational support;
and \S4 discusses a more realistic description in which the halo possesses
sub-structure.

\section{Spherical halo}
The simplest Galactic halo model is the isothermal sphere. In the context of
a cold cloud dark matter picture, this model gives some simple results against
which one can usefully compare the data.

\subsection{Atomic gas from molecular clouds}
Sound speeds for the gas within the clouds are expected to be very low
-- $0.2\;{\rm km\,s^{-1}}$ in the model of Wardle \& Walker (1998; ApJL
in preparation) -- in comparison
with the relative speeds of clouds within the halo (several hundred km/s),
so the most basic expectation is that strong shocks will be generated during
collisions and these will unbind the gas. Providing only that the shock speeds
exceed $24\;{\rm km\,s^{-1}}$ (Kwan 1977), the molecular gas in the clouds will
be completely dissociated during a head-on collision.\footnote{Dwarf galaxies with
internal velocity dispersions $\lta17\;{\rm km\,s^{-1}}$ should consequently possess
interstellar media which are predominantly molecular.} At higher speeds the
ionised fraction increases (50\% H ionisation at $\simeq70\;{\rm km\,s^{-1}}$;
Shull \& McKee 1979), being unity in the case of the Galactic halo (shock speeds
$\simeq200\;{\rm km\,s^{-1}}$). In the dense ($\sim10^{12}\;{\rm cm^{-3}}$),
post-shock gas, radiative cooling proceeds more rapidly than re-expansion,
implying that most of the dissipated kinetic energy is {\it radiated\/} away
-- the ramifications of this are discussed elsewhere (Walker \& Wardle, 1998c,
in preparation). Thermal instability is present in the post-shock gas, resulting
in hot, low-density gas in which cooler, denser blobs are embedded. Recombination
of the ions is also very rapid, but molecular re-formation is relatively slow, and
consequently, {\it atomic\/} gas is released into the Galactic halo in the case of
head-on cloud-cloud collisions.

In the case of glancing collisions, a good part of the mass of the clouds escapes
unshocked, and is not significantly decelerated. However, some fraction of the
power radiated by the shocked gas will be absorbed by the unshocked material and,
because the total radiated energy is many orders of magnitude larger than the
clouds' thermal energy, it is expected that complete unbinding will still occur.
What is less certain is the material state of the unshocked gas, which could
remain molecular. (Note, however, that shocks caused by subsequent interaction
with the surrounding medium should even then convert the gas to atomic form.) In the
following discussion we shall restrict attention to head-on collisions, and the
subsequent fate of the gas released by such collisions.

\subsection{Expansion and merging}
The radiative phase of each collision reduces the temperature from
$T\simeq5\times10^5$~K to $T\simeq6000$~K, at which point the
radiation time-scale exceeds the expansion time-scale and subsequent evolution is
approximately adiabatic. During this phase the gas expands at roughly
$15\;{\rm km\,s^{-1}}$ from $R\sim10^{14}$~cm  to $R\sim10^{18}$~cm; for most of
this period the gas is so cold that it could be detected only in absorption. The
expansion ceases to be adiabatic when the density has dropped to the point where
interaction with the external medium becomes important. For an external density
$\sim10^{-2}\;{\rm cm^{-3}}$ (see \S\S2.4,2.5) this occurs roughly
$10^4$~yr after the collision. At this point the material is re-heated to
several thousand degrees by the passage of a shock, rendering it visible once
more in emission. Subsequently -- on a time-scale of order $10^5$~yr -- the
post-collision gas loses its kinematic identity as its velocity approaches
that of the surrounding fluid.

\subsection{Velocity field}
As the constituents of a spherical halo have zero mean velocity field, predicting
the {\it native\/} velocity field of post-collision gas is trivial: as seen from the
LSR it is just $-V\sin l\,\cos b$, where $V=220\;{\rm km\,s^{-1}}$ is the circular
speed of the LSR. Around this mean we expect a dispersion of $55\;{\rm km\,s^{-1}}$,
much smaller than for the pre-collision clouds ($155\;{\rm km\,s^{-1}}$). This is a
direct consequence of the fact that the collision process favours pairs of clouds
having antiparallel velocities. Note that for as long as the gas remains kinematically
distinct (separated in velocity by at least $100\;{\rm km\,s^{-1}}$) from its
surroundings, there is negligible acceleration in the Galactic potential, simply
because this phase lasts for much less than an orbital period --- $10^5$~yr versus
$3\times10^8$~yr. 

\subsection{Detectability of post-collision clouds}
A key point to notice is that the low mass of the colliding clouds
$\sim10^{-3}\;{\rm M_\odot}$ renders the collision products individually
undetectable unless they are very close to the Sun. Distances $\lta70$~pc
are required for the collision products ($\sim2\times10^{-3}\;{\rm M_\odot}$)
to be detectable by HIPASS (Staveley-Smith 1997), placing these clouds within the
Local Hot Bubble. If the LHB density is $10^{-2}\;{\rm cm^{-3}}$
(Cox \& Reynolds 1987), each expanding cloud will approach the LSR velocity on
a time-scale $\sim10^5$~yr. This implies median angular radii of order
$1\;M_{-3}^{-1/6}$~deg. From the cloud-cloud collision rate per unit volume
(Walker 1998), we compute an expected HIPASS detection rate
of order $1\,M_{-3}^{5/6}\;{\rm sr^{-1}}$, where the mass of each cold cloud
is written as $10^{-3}M_{-3}\;{\rm M_\odot}$.

The small number of individually detectable post-collision clouds indicates
that this model is germane to only a fraction of the observed population.
Moreover, some of the large HVC complexes now have distance constraints
(van~Woerden 1998) that exclude nearby, low-mass clouds. By the same token, though,
at least one HVC of sub-solar mass is known (van Woerden 1998), so the model may
indeed be relevant to a subset of the observed HVC population.

\subsection{21~cm background}
As noted above, the expanding gas from individual collisions is not detectable
unless the collision occurred very local to the Sun (roughly, within the Local
Hot Bubble). As the halo extends well beyond this volume, we see that the bulk
of the post-collision population should manifest itself as an unresolved background.
This population is born kinematically cold (velocity dispersion $55\;{\rm km\,s^{-1}}$),
and therefore inadequately supported in the Galactic potential. Infall will occur,
presumably at a speed comparable to the typical sound speed in the flow. We
estimate that this is $\sim10\;{\rm km\,s^{-1}}$ -- neutral gas at several
thousand Kelvin -- implying an infall time of order $4\times10^8$~yr. Shock
heating will create localised regions which are hotter, and may be highly ionised,
but cooling and recombination are, at a density $\sim10^{-2}\;{\rm cm^{-3}}$, relatively
rapid. Given the cloud-cloud collision rate per unit volume (Walker 1998) it
is straightforward to deduce the rate per unit area towards the Galactic poles,
say. Knowing the infall time-scale then allows us to deduce a high-latitude column
of $N_H\sim10^{20}\;{\rm cm^{-2}}$ in warm, predominantly neutral gas.

We are unable to give quantitative predictions for the velocity field of this
fluid as a whole; qualitatively we can say that the gas should be sub-rotating
(relative to the disk) and infalling. Turbulence is expected to be present as a
consequence of the ongoing injection (with velocity dispersion $55\;{\rm km\,s^{-1}}$)
of fresh gas, from recent collisions. Shocks will tend to damp any turbulence; on
the other hand, kinetic energy liberated by infall will tend to sustain it.

Despite the uncertainties, the characteristics we anticipate for the infalling
halo gas are somewhat reminiscent of the large velocity dispersion component
($\sigma_v\simeq60\;{\rm km\,s^{-1}}$, $N_H=1.4\times10^{19}\;{\rm cm^{-2}}$)
discovered in the 21-cm Leiden-Dwingeloo Survey (Kalberla et al 1998). We caution
that such a large velocity dispersion component has not been identified in previous
HI surveys (Dickey \& Lockman 1990).

\section{Rotating halo}
Given that the visible disk of our Galaxy is almost entirely supported by
its rotation, and in our model this material was originally part of the
dark halo, it is important to consider halo rotation.  Analytic models
of rotating isothermal halos have been given by Toomre (1982). However,
unless a large amount of rotation is invoked -- a possibility which would
be difficult to understand in the context of a tidal torqueing origin for
halo angular momentum -- the discussion in the previous section is only changed
slightly. The most significant change is that the apparent native velocity field
of post-collision clouds becomes $[\langle v_\phi\rangle-V]\,\sin l\,\cos b$,
with a dispersion of roughly $55\;{\rm km\,s^{-1}}$. In principle, comparison
with an observed velocity field could admit a measure of the halo rotation
speed, $\langle v_\phi\rangle$, but interaction with low velocity disk gas
renders this a difficult procedure. 

\section{Structured halos}
Even if one does not entirely subscribe to the current paradigm of
hierarchichal formation of structure in the Universe (see, e.g., Peebles 1993),
one still expects galaxy halos, including the Galactic halo, to
incorporate substructure, as a consequence of the accretion of smaller
satellite galaxies. This modifies the picture of \S\S2,3 substantially by
introducing both density and velocity structure, and we can no-longer
give unique predictions because the nature of these structures is not known
{\it a priori.\/} Nevertheless we can outline some general features
as follows.

\begin{figure}[t]
\plotone{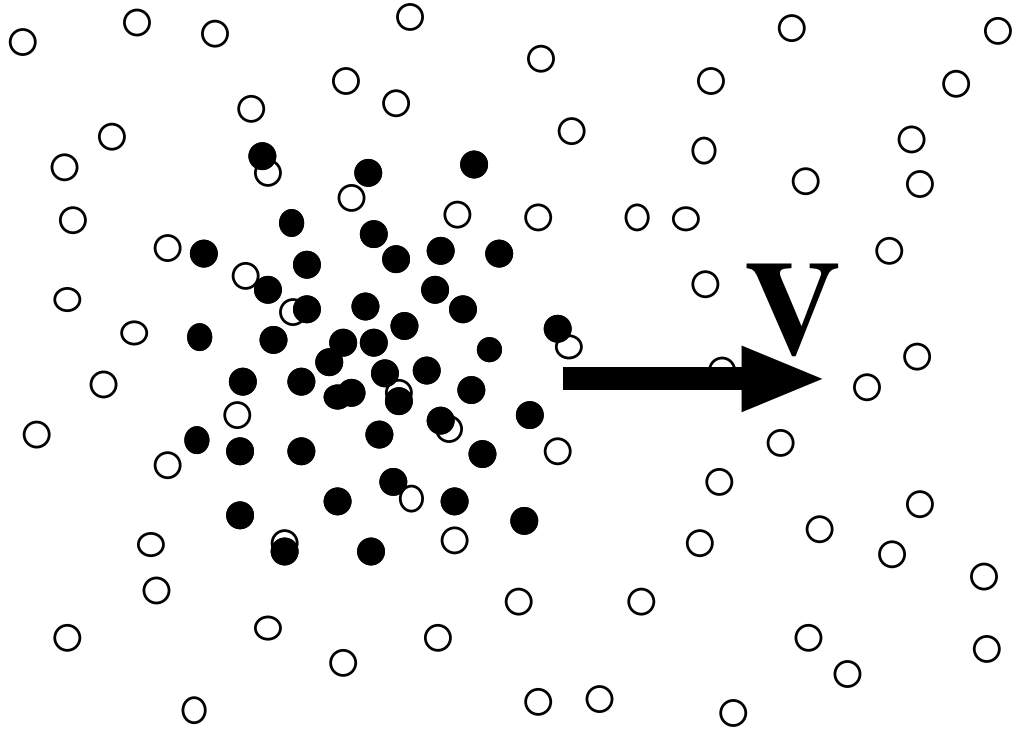}
\caption{A satellite composed of cold clouds (filled circles)
moving through a background of similar clouds (open circles) comprising
the quasi-spherical component of the Galactic dark halo. Collisions
occurring between these two sets of clouds lead to a wake of diffuse
atomic gas behind the satellite.}
\end{figure}
For any given satellite we can imagine the dark matter as following a
single orbit, with a relatively small space/velocity dispersion, at an orbital
speed of order the Galaxy's circular speed ($V$) --- as per figure 1. Associated
with this satellite there are two types of cloud-cloud collisions to
contemplate: those where clouds within the satellite collide with each
other, and those where one of the colliding pair is part of a kinematically
separate component of the Galaxy's halo. In particular we shall assume this
second component to be a quasi-spherical background of dark matter, as per
the model halo considered in \S2. For collisions occurring between clouds
within the satellite, the post-collision gas remains bound to the satellite
(though it may subsequently be stripped out by ram pressure). If this process
dominates it presumably leads to a recognisable satellite {\it galaxy,\/} with
its own stars and interstellar medium.

For a satellite interior to the dark halo core radius of the Galaxy -- i.e.
about 6~kpc (Walker 1998) -- the dominant process is expected
to involve ``satellite clouds'' colliding with ``spherical halo clouds.''
The centre-of-mass velocity for this type of collision differs by
$\sim V/2\sim100\;{\rm km\,s^{-1}}$ from the satellite's orbit, and is thus
very unlikely to be bound to it. (Notice, also, the trivial point that a tidally
disrupted satellite halo -- i.e. a stream of dark matter -- cannot bind gas
released in {\it any\/} type of collision.) In consequence we expect satellite
halos to leave behind them a trail of gas.

The details of the collision process itself, and subsequent adiabatic
phase, do not differ sufficiently from the case of the spherical halo
(\S\S2.1,2.2) that they warrant separate discussion. The merging process,
on the other hand, is slightly different. A qualitative distinction
is that there is no ongoing injection of gas (hence kinetic energy) into the
wake downstream from the satellite: there is only ``fossil turbulence'',
which is damped by conversion into thermal energy. Quantitatively we note that
the velocity dispersion of the merging gas is larger --- up to $80\;{\rm km\,s^{-1}}$.
This is because the satellite halo has a velocity dispersion that is {\it small\/}
in comparison to that of the Galactic halo, so the velocities of the colliding cold
clouds do not cancel as effectively as in the case of collisions within a simple
spherical halo. As a result the temperatures reached by gas in the wake may
initially be as high as $10^5$~K, but radiative cooling rapidly brings this down
to values less than $10^4$~K, and the gas will recombine on a time-scale $\sim10^7$~yr.

A purely dark satellite of mass $M_{sat}$ moving on an approximately circular
orbit around the Galaxy will leave behind gas at the rate $-\dot M_{sat}
\simeq 8\rho V M_{sat}/\Sigma$, where $\rho$ is the density of the spherical
component of the Galaxy's dark halo, and $\Sigma\simeq130\;{\rm g\,cm^{-2}}$
(Walker 1998) is the mean surface density of the individual dark clouds. (Note
that the collision cross-section for pairs of identical clouds is four times the
geometric cross-section of a single cloud, because even glancing collisions
will unbind the gas.) For a satellite orbiting near the solar circle, we
then expect that between successive crossings of the Galactic plane a mass
of roughly $3\times10^{-3}M_{sat}$ will be deposited in the wake.

Evidently it is possible to explain the masses, at least, of the large complexes
of HVCs (van~Woerden 1998) if one contemplates satellite dark halos as massive
as $10^9\;{\rm M_\odot}$. Because satellite orbits can be retrograde, with
respect to the Galactic disk, these wakes may display LSR velocities of magnitude
$\sim2V\sim400\;{\rm km\,s^{-1}}$, and may thus be relevant to the Very-HVCs
(Wakker \& van Woerden 1997). Of course,
the trajectory of the gas wake differs substantially from that of the satellite
itself. Crudely speaking, gas is injected into the wake with a mean velocity
parallel to, but only half as large as the satellite's velocity; acceleration in
the Galactic potential subsequently pulls the wake progressively further away from
the satellite's trajectory as it progresses downstream. Consequently these
wakes are expected to display a substantial degree of infall in their kinematics. 

\section{Summary}
The combined model of a quasi-spherical Galactic halo (\S\S2,3) with satellite
substructure (\S4), and cold clouds making up the dark matter, gives a rich
theoretical picture of high velocity atomic gas arising from cloud-cloud
collisions. The quasi-spherical component leads one to expect a small number
of low mass HVCs located in the Local Hot Bubble, and conforming to a simple
velocity field. These HVCs are just the nearest examples of a much larger
population, which gives rise to a background of broad 21~cm emission.
Generalising to a model which includes satellite substructure within the
dark halo leads us to expect gaseous wakes containing large quantities of
high velocity atomic gas. These wakes may be relevant to the observed complexes
and streams of HVCs, including the Magellanic Stream. We emphasise, however,
that a satellite {\it galaxy\/} -- i.e. including stars -- is not required
as the wake is produced by dark matter.

This paper presents some ideas on how the cold-cloud picture of dark matter
may help us to understand the presence of high velocity gas in our Galaxy; but
it is interesting to ask ``What do HVCs tell us about dark matter?'' (In the
context of our model, of course!) Obviously one could learn a great deal about
the structure of the dark halo of our Galaxy, and in turn this would inform
us about halo formation and so on. However, the most significant point at
present seems to be the fact that HVCs contain metals, while the model dictates
that they should be composed of primordial gas. One could take the view that
this rules out the model, if one had complete faith in the conventional
perspective on Big Bang nucleosynthesis. Alternatively one could argue that
the gas has been ``polluted'' by a population of massive stars in the early
Universe; or perhaps that the metals actually come from diffuse Galactic
gas swept up by the expanding post-collision material? Nevertheless we caution
that the metallicity of primordial gas must, ultimately, be determined by
observation.

\section{Acknowledgments}
The Special Research Centre for Theoretical Astrophysics is funded by the
Australian Research Council under its Special Research Centres program.
MW enjoyed the Stromlo HVC workshop.

\end{document}